# Emotion-Aware Interaction Design in Intelligent User Interface Using Multi-Modal Deep Learning

Shiyu Duan
Carnegie Mellon University
Pittsburgh, USA

Ziyi Wang
University of Maryland
College Park, USA

Shixiao Wang
School of Visual Arts
New York, USA

Mengmeng Chen
New York University
New York, USA

Runsheng Zhang
University of Southern California
Los Angeles, USA

*Abstract*— **In an era where user interaction with technology is ubiquitous, the importance of user interface (UI) design cannot be overstated. A well-designed UI not only enhances usability but also fosters more natural, intuitive, and emotionally engaging experiences, making technology more accessible and impactful in everyday life. This research addresses this growing need by introducing an advanced emotion recognition system to significantly improve the emotional responsiveness of UI. By integrating facial expressions, speech, and textual data through a multi-branch Transformer model, the system interprets complex emotional cues in real-time, enabling UIs to interact more empathetically and effectively with users. Using the public MELD dataset for validation, our model demonstrates substantial improvements in emotion recognition accuracy and F1 scores, outperforming traditional methods. These findings underscore the critical role that sophisticated emotion recognition plays in the evolution of UIs, making technology more attuned to user needs and emotions. This study highlights how enhanced emotional intelligence in UIs is not only about technical innovation but also about fostering deeper, more meaningful connections between users and the digital world, ultimately shaping how people interact with technology in their daily lives.**

## I. INTRODUCTION

With the rapid development of technology, the design and development of intelligent user interfaces (UI) plays an increasingly important role in the field of modern technology. Traditional UI mainly relies on user input, such as button clicks or sliding operations for interaction [1]. However, with the continuous advancement of artificial intelligence (AI) and deep learning technology [2-4], intelligent UI has gradually integrated emotion recognition functions to provide a more personalized and adaptive interactive experience through the perception of the user's emotional state. This kind of intelligent UI based on emotion recognition can be applied in a variety of scenarios, including risk alert [5-8], health monitoring [9-12], computer vision [13], customer service, and other fields, effectively improving the quality and efficiency of user experience.

Emotion recognition technology relies on the analysis of multi-modal data such as users' facial expressions, voice characteristics, body movements, and text information to determine the user's current emotional state [14]. In recent years, the successful application of deep learning, especially Transformer-based models, in the fields of natural language processing (NLP) [15] and computer vision [16] has significantly improved the performance of emotion recognition systems. The advantage of the Transformer model is that it can capture the long-range dependencies of data and effectively focus on the most important information through the self-attention mechanism, thereby providing more accurate analysis in the emotion recognition process. By integrating Transformer emotion recognition technology, the intelligent UI can dynamically adjust the interface layout, interaction methods, content recommendations, etc. to better meet the user's emotional needs and preferences, thereby achieving a more natural interaction effect.

In the background, research on emotion recognition technology has broad practical significance. First of all, in the field of health monitoring, intelligent UI based on emotion recognition can monitor the user's emotional fluctuations in real-time and intervene when the user experiences negative emotions or stress, such as automatically recommending soothing music, relaxation training videos, etc [17]. Secondly, in the field of education, intelligent UI can determine the emotional state of students during the learning process, such as anxiety, fatigue, or distraction, so as to automatically adjust the learning content or recommend a break. Finally, in the field of customer service, intelligent UI can determine whether the customer is satisfied by analyzing the customer's emotional response during the service process, and then adjusting the service strategy or recommending different products to improve customer satisfaction and loyalty.

For this research topic, we designed an emotion recognition algorithm based on Transformer and applied it to the interaction design of intelligent UI. Our algorithm combines self-attention mechanism and multi-head attention mechanism to handle multi-modal emotional data. Specifically, we adopt a multi-modal Transformer architecture, with one branch used to process facial image sequences and another branch used to analyze the time series characteristics of speech data. In addition, we also added a text analysis branch to process users' text input, such as chat logs or user comments. Through the multi-modal fusion mechanism, the model can synthesize information from multiple modalities, thereby achieving higher accuracy and stronger robustness in the emotion classification process.

In practical applications, this emotion recognition algorithm can be integrated into intelligent UI to achieve the ability to dynamically adjust interface content. For example, when the system detects that a user is showing boredom or disinterest while watching a video, it can automatically switch to the user's more preferred content type. In addition, in games, intelligent UI can determine the player's experience in the game based on changes in their expressions, such as happiness, excitement, or frustration, thereby dynamically adjusting the difficulty of the game or providing incentive mechanisms to enhance the user's immersion and participation. Such an intelligent emotional response mechanism not only improves the level of personalization of user experience, but also increases the frequency and effectiveness of interaction between the system and users.

To sum up, the application of emotion recognition based on intelligent UI has broad prospects. By sensing and responding to user emotions in real-time, smart UI can provide a more natural and efficient human-computer interaction experience. In the future, we will continue to optimize the emotion recognition algorithm, improve its accuracy and robustness in different scenarios, and explore more application scenarios, such as smart homes, online education, and virtual reality. Through continuous innovation and research, we believe that intelligent UI based on emotion recognition will bring more convenient and pleasant experiences to people's lives.

## II. RELATED WORK

In recent years, multi-modal deep learning and Transformer-based architectures have significantly influenced the development of emotion-aware interaction in intelligent user interfaces (UIs). These methodologies enable UIs to recognize and respond to users' emotions, enhancing personalized and empathetic interaction. Yan et al. [18] contributed to the interpretation of multi-dimensional time series data by transforming them into interpretable event sequences, providing a foundation for multi-modal data processing crucial for real-time emotion recognition. This approach is particularly relevant to emotion-aware UIs where processing diverse data streams, such as facial expressions and vocal cues, requires models capable of handling complex and synchronized time-based inputs effectively.

In The robustness of emotion recognition models in dynamic environments can benefit from Jiang et al.'s [19] work on cross-domain adaptability through adversarial networks, which stabilize the model across varied data inputs. Similarly, Du et al. [20] addressed the challenges of semantic complexity with Transformer models in language processing, informing the nuanced textual interpretation required in emotion-aware UIs for detecting emotional states accurately. Optimizing model efficiency, Chen et al. [21] introduced retrieval-augmented generation to enhance response relevance in real-time applications—a strategy adaptable to emotion-aware UIs needing rapid contextual responses. Furthermore, Huang et al. [22] demonstrated knowledge distillation for efficient model performance, a technique that could support real-time emotion recognition by streamlining the model's structure.

Graph neural networks (GNNs) were leveraged by Wei et al. [23] for enhanced feature extraction across diverse data, relevant for synthesizing text sentiment, vocal tone, and facial expressions within a multi-modal UI framework. In alignment with adaptive UI responses, Wang et al. [24] explored large language models (LLMs) for automated, genre-aware feedback, underscoring the value of dynamic response generation. Lastly, Liu et al. [25] explored self-attention and embedding for personalized recommendations, aligning with the goal of adapting UI responses to users' emotional states in real time.

## III. METHOD

In this study, we proposed a multimodal emotion recognition method based on the Transformer architecture. In addition to its emotion recognition capabilities, our intelligent UI dynamically adjusts its content based on the user's emotional state. For example, when boredom is detected, the system might switch to a more interactive mode by introducing engaging visual content or recommending interactive quizzes. When positive emotions like excitement are detected, the UI can intensify engagement by providing additional features or gamified challenges. These adaptive content strategies are vital to maintaining user engagement and ensuring a personalized experience. First, we input multimodal data, including facial image sequences, speech signals, and text data, into their respective Transformer branches. The Transformer module of each branch consists of multiple self-attention layers and feed-forward neural networks to capture the feature representation of the input data. The overall architecture of the transformer is shown in Figure 1.

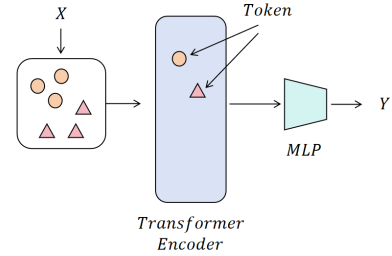

Figure 1 Transformer overall architecture

For facial image sequences, we represent the image sequence as a vector $X_{img} \in R^{T \times d_{img}}$, where T represents the sequence length and $d_{img}$ represents the dimension of each frame image feature. We calculate the importance of the feature vector at each time step through the self-attention mechanism, as follows:

$$Zimg = \text{Soft}\max(\frac{QK^T}{\sqrt{d_k}})V$$

Where Q, K, and V are query, key, and value matrices obtained by linearly transforming the input image feature vector, and $\sqrt{d_k}$ is the dimension of the key. This mechanism enables the model to capture global information throughout the entire image sequence, thereby better understanding the dynamic changes of emotions.

For speech data, we process it into a time series vector $X_{audio} \in R^{L \times d_{audio}}$, where L represents the sequence length and $d_{audio}$ is the feature dimension for each time step. Similar to the processing of image sequences, we apply a self-attention mechanism on speech data to enable the model to identify emotion-related speech features. By calculating the attention weights, we obtain a context feature vector representing the speech sequence.

The text data is also input in sequence form, which we represent as $X_{text} \in R^{N \times d_{text}}$, where N is the length of the text sequence and $d_{text}$ is the embedding dimension of each word. We use a pre-trained language model (such as BERT) for preliminary encoding, and then input the encoded features into the Transformer module to further extract sentiment-related features.

Next, we fuse the feature vectors $Z_{img}$, $Z_{audio}$, and $Z_{text}$ from three different modalities. In order to achieve effective fusion of multimodal features, we designed a weighted fusion mechanism:

$$Z_{fused} = \alpha Z_{img} + \beta Z_{audio} + \chi Z_{text}$$

$\alpha$, $\beta$, and $\chi$ are the importance weights of each modal feature, which are obtained through adaptive learning. In this way, we can ensure that the model can automatically adjust the influence of each modality during the fusion process to improve the accuracy of sentiment classification.

Finally, we pass the fused feature vector through the fully connected layer and the Softmax layer to output the final sentiment classification result. The loss function uses the cross-entropy loss:

$$L = -\sum_{c=1}^{C} y_c \log(y'_c)$$

Where C is the number of emotion categories, $y_c$ is the actual label, and $y'_c$ is the probability value predicted by the model. We optimize the model parameters by minimizing the loss function so that it can more accurately identify emotion categories.

In order to further improve the performance of the model, we used data augmentation technology during the training process so that the emotion recognition system can still perform well with limited data. The overall method shows high accuracy and generalization ability in multimodal emotion recognition.

IV. EXPERIMENT

*A. Datasets*

In this study, we used the MELD (Multimodal EmotionLines Dataset) dataset for model training and verification. The MELD dataset is a public multimodal emotion recognition dataset designed specifically for studying emotion recognition in conversation scenarios. It comes from a conversation segment of the TV series "Friends". Each sample contains a facial expression image sequence, a corresponding voice signal, and a text conversation content, covering seven major emotion categories: joy, anger, sadness, fear, surprise, disgust, and neutral.

The uniqueness of the MELD dataset is that it not only contains the emotion label of each interlocutor, but also takes into account the conversation context information, allowing researchers to use the context to more accurately predict emotions when training the model. In addition, each sample in the dataset is multimodal, including a facial image sequence (showing the facial changes of the interlocutor during the conversation), a voice file (capturing the speaker's tone and acoustic characteristics), and the corresponding text (i.e., the conversation content). The combination of these multimodal data provides comprehensive information for the model, making the emotion recognition process more accurate and multi-dimensional.

By using the MELD dataset, we were able to fuse visual, speech, and text information in the emotion recognition model, fully train the Transformer-based model, and test its performance in the multimodal emotion recognition task. The richness and diversity of the dataset ensures that the model can learn and perform accurate emotion classification in a variety of scenarios and emotions, thus verifying the effectiveness of our approach.

*B. Experiments*

In the experiment, we selected five comparative experimental models to evaluate the performance of the emotion recognition system. These models are all deep learning models, including: 1) Emotion recognition model based on unimodal Transformer (using only text); 2) Multimodal model combining CNN and LSTM (processing facial images and speech); 3) Emotion recognition model integrating BERT and LSTM (processing text and speech); 4) Using an improved multimodal Transformer architecture, combining facial expressions, speech and text (MM-Transformer); 5) The weighted Transformer model of full modal fusion proposed by us. Each model was trained and tested on the same experimental environment and dataset, and the evaluation indicators were accuracy (Acc) and F1 value. The following are the experimental results:

Table 1 Experiment result

| Model | Auc | F1 |
|---|---|---|
| Transformer | 0.685 | 0.653 |
| CNN-LSTM | 0.724 | 0.698 |
| BERT-LSTM | 0.756 | 0.727 |
| MM-Transformer | 0.789 | 0.761 |
| Ours | 0.817 | 0.795 |

From the experimental results, it can be seen that the performance of different models in emotion recognition tasks gradually improves with the complexity of the architecture and the degree of fusion of multimodal features. First, although the unimodal Transformer model can perform well in text feature extraction, due to the lack of data support from other modalities such as vision and speech, its AUC and F1 values are only 0.685 and 0.653 respectively. This shows that single-modal information cannot fully capture the user's emotional changes and complex emotional characteristics in emotion recognition tasks. In contrast, the CNN-LSTM model combines image and speech data, making the extraction of emotional features richer, thereby improving the overall performance of the model, with AUC and F1 values reaching 0.724 and 0.698 respectively. This shows that multimodal data fusion helps to improve the accuracy of emotion recognition and enhances the model's ability to distinguish different emotion categories.

With the continuous upgrading of the model architecture, the BERT-LSTM model has shown higher emotion recognition capabilities, with an AUC value of 0.756 and an F1 value of 0.727. This result verifies the strong advantages of BERT in text processing and the high efficiency of LSTM in processing time series data. The combination of the two enables the model to capture emotional features more comprehensively. The MM-Transformer model that follows closely uses the self-attention mechanism in the process of multimodal data fusion, and further improves the performance of emotion recognition by selectively focusing on information of different modalities. Its AUC and F1 values are 0.789 and 0.761 respectively, indicating that in the task of emotion recognition, the use of the Transformer architecture to adaptively integrate multimodal information can effectively improve the accuracy and robustness of the model.

In the end, the model we proposed performed best in the experiment, achieving an AUC of 0.817 and an F1 value of 0.795. This result shows that through the efficient Transformer architecture and the deep fusion of multimodal information, our method still shows higher robustness and accuracy in emotion recognition. Compared with other models, our method shows better feature extraction and fusion capabilities when processing multimodal data, effectively improving the accuracy and stability of emotion recognition. These results prove that in emotion recognition tasks, making full use of multimodal data and adopting a reasonable deep learning architecture design are the keys to improving the performance of emotion recognition systems. Finally, we also give the rising graph of AUC during training, the result is shown in Figure 2.

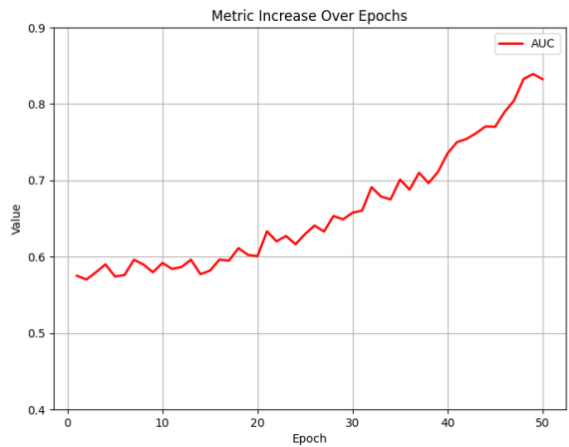

Figure 2 AUC result rise chart

V. CONCLUSION

This research substantiates the pivotal role of user interfaces (UIs) in mediating the interaction between technology and users by implementing a multi-modal emotion recognition method based on Transformer architecture. The integration of facial imagery, vocal expressions, and textual data enables our system to capture a holistic view of user emotions, significantly enhancing the emotional intelligence and classification accuracy of UIs. Our empirical investigations, utilizing the public MELD dataset alongside a series of comparative experiments, have affirmed the efficiency and superiority of our multi-modal data fusion approach and Transformer architecture. Notably, enhancements in the model's complexity correlate with improved performance in emotion recognition tasks, illustrating the critical importance of comprehensive multi-modal data utilization for advancing emotional analytical precision in UIs. The proposed emotion-aware UI significantly enhances the user experience by making technology interactions more intuitive and empathetic. For instance, the system can recognize when a user is disengaged and adapt by recommending alternative content. This dynamic adaptability creates a more engaging experience that aligns more closely with user needs. Unlike traditional UIs, which lack emotional awareness, this system provides a unique empathy-driven approach that fosters deeper and more meaningful user engagement. The significance of UIs extends beyond mere functionality; they are essential for enabling intuitive, seamless, and empathetic interactions that align with the evolving expectations of digital users. Emotionally intelligent UIs enhance user satisfaction by providing more personalized and responsive interactions, which are crucial for fostering deeper user engagement and loyalty. This study not only contributes to the theoretical advancements in UI design but also underscores the practical implications of emotionally aware interfaces in enhancing user experience. Despite the notable achievements of this research, the acquisition and processing of multi-modal data present ongoing challenges, particularly in practical applications. To overcome these limitations and ensure the model's applicability across diverse scenarios, future initiatives will focus on integrating transfer learning and data augmentation strategies to bolster the model's robustness and generalization capabilities. In pursuit of refining human-computer interaction, this research reaffirms the necessity of continuous technological enhancements and application diversification. As UIs evolve to become more adept at interpreting and responding to human emotions, they promise to redefine the paradigms of user interaction with technology, emphasizing the critical role of UIs in the landscape of digital innovation. This commitment to developing highly sophisticated, emotionally intelligent UIs is fundamental to realizing more natural, insightful, and user-centric technological interactions.

References

[1] M. R. Fachrizal, A. Paramitha Fadillah and A. Budiarto, "UI/UX Prototype Usability Analysis of E-Commerce Websites," *2023 International Conference on Informatics Engineering, Science & Technology (INCITEST)*, pp. 1-6, 2023

[2] J. Hu, Y. Cang, G. Liu, M. Wang, W. He, and R. Bao, "Deep Learning for Medical Text Processing: BERT Model Fine-Tuning and Comparative Study," arXiv preprint arXiv:2410.20792, 2024.

[3] M. Sui, J. Hu, T. Zhou, Z. Liu, L. Wen, and J. Du, "Deep Learning-Based Channel Squeeze U-Structure for Lung Nodule Detection and Segmentation," arXiv preprint arXiv:2409.13868, 2024.

[4] Z. Xu, J. Pan, S. Han, H. Ouyang, Y. Chen, and M. Jiang, "Predicting Liquidity Coverage Ratio with Gated Recurrent Units: A Deep Learning Model for Risk Management," arXiv preprint arXiv:2410.19211, 2024.

[5] K. Xu, Y. Wu, H. Xia, N. Sang, and B. Wang, "Graph Neural Networks in Financial Markets: Modeling Volatility and Assessing Value-at-Risk," Journal of Computer Technology and Software, vol. 1, no. 2, 2022.

[6] B. Liu, I. Li, J. Yao, Y. Chen, G. Huang, and J. Wang, "Unveiling the Potential of Graph Neural Networks in SME Credit Risk Assessment," arXiv preprint arXiv:2409.17909, 2024.

[7] M. Sun, W. Sun, Y. Sun, S. Liu, M. Jiang, and Z. Xu, "Applying Hybrid Graph Neural Networks to Strengthen Credit Risk Analysis," arXiv preprint arXiv:2410.04283, 2024.

[8] Y. Dong, J. Yao, J. Wang, Y. Liang, S. Liao, and M. Xiao, "Dynamic Fraud Detection: Integrating Reinforcement Learning into Graph Neural Networks," Proceedings of the 2024 6th International Conference on Data-driven Optimization of Complex Systems (DOCS), pp. 818-823, 2024.

[9] Y. Zi, X. Cheng, T. Mei, Q. Wang, Z. Gao, and H. Yang, "Research on Intelligent System of Medical Image Recognition and Disease Diagnosis Based on Big Data," Proceedings of the 2024 IEEE 2nd International Conference on Image Processing and Computer Applications (ICIPCA), pp. 825-830, 2024.

[10] W. He, R. Bao, Y. Cang, J. Wei, Y. Zhang, and J. Hu, "Axial attention transformer networks: A new frontier in breast cancer detection," arXiv preprint arXiv:2409.12347, 2024.

[11] Y. Li, W. Zhao, B. Dang, X. Yan, M. Gao, W. Wang, and M. Xiao, "Research on adverse drug reaction prediction model combining knowledge graph embedding and deep learning", Proceedings of the 2024 4th International Conference on Machine Learning and Intelligent Systems Engineering (MLISE), pp. 322-329, June 2024.

[12] D. Sun, M. Sui, Y. Liang, J. Hu, and J. Du, "Medical Image Segmentation with Bilateral Spatial Attention and Transfer Learning," Journal of Computer Science and Software Applications, vol. 4, no. 6, pp. 19-27, 2024.

[13] Z. Wu, H. Gong, J. Chen, Z. Yuru, L. Tan, and G. Shi, "A Lightweight GAN-Based Image Fusion Algorithm for Visible and Infrared Images," arXiv preprint arXiv:2409.15332, 2024.

[14] J. Du, Y. Cang, T. Zhou, J. Hu, and W. He, "Deep Learning with HM-VGG: AI Strategies for Multi-modal Image Analysis," arXiv preprint arXiv:2410.24046, 2024.

[15] S. Liu, G. Liu, B. Zhu, Y. Luo, L. Wu, and R. Wang, "Balancing Innovation and Privacy: Data Security Strategies in Natural Language Processing Applications," arXiv preprint arXiv:2410.08553, 2024.

[16] H. Liu, B. Zhang, Y. Xiang, Y. Hu, A. Shen, and Y. Lin, "Adversarial Neural Networks in Medical Imaging Advancements and Challenges in Semantic Segmentation," arXiv preprint arXiv:2410.13099, 2024.

[17] J. J. Sundi, H. Kumar, and R. Bedi, "Real-Time Facial Expression Recognition Using Convolutional Neural Networks for Adaptive User Interfaces," *Proceedings of the 2024 5th International Conference for Emerging Technology (INCET)*, IEEE, pp. 1-6, 2024.

[18] X. Yan, Y. Jiang, W. Liu, D. Yi, and J. Wei, "Transforming Multidimensional Time Series into Interpretable Event Sequences for Advanced Data Mining," arXiv preprint, arXiv:2409.14327, 2024.

[19] M. Jiang, J. Lin, H. Ouyang, J. Pan, S. Han, and B. Liu, "Wasserstein Distance-Weighted Adversarial Network for Cross-Domain Credit Risk Assessment," arXiv preprint, arXiv:2409.18544, 2024.

[20] J. Du, Y. Jiang, and Y. Liang, "Transformers in Opinion Mining: Addressing Semantic Complexity and Model Challenges in NLP," Transactions on Computational and Scientific Methods, vol. 4, no. 10, 2024.

[21] J. Chen, R. Bao, H. Zheng, Z. Qi, J. Wei, and J. Hu, "Optimizing Retrieval-Augmented Generation with Elasticsearch for Enhanced Question-Answering Systems," arXiv preprint arXiv:2410.14167, 2024.

[22] G. Huang, A. Shen, Y. Hu, J. Du, J. Hu, and Y. Liang, "Optimizing YOLOv5s Object Detection through Knowledge Distillation Algorithm," arXiv preprint arXiv:2410.12259, 2024.

[23] J. Wei, Y. Liu, X. Huang, X. Zhang, W. Liu, and X. Yan, "Self-Supervised Graph Neural Networks for Enhanced Feature Extraction in Heterogeneous Information Networks," arXiv preprint, arXiv:2410.17617, 2024.

[24] C. Wang, Y. Dong, Z. Zhang, R. Wang, S. Wang, and J. Chen, "Automated Genre-Aware Article Scoring and Feedback Using Large Language Models," arXiv preprint arXiv:2410.14165, 2024.

[25] W. Liu, R. Wang, Y. Luo, J. Wei, Z. Zhao, and J. Huang, "A Recommendation Model Utilizing Separation Embedding and Self-Attention for Feature Mining," arXiv preprint, arXiv:2410.15026, 2024.